\documentclass[]{pasj02} 
\usepackage{url}
\usepackage{lscape}

\jyear{2026}
\Received{}
\Accepted{}

\begin{document} 

\title{ Opposition effect of comet 28P/Neujmin observed with Subaru Hyper Suprime-Cam }

\author{
 Takafumi \textsc{Ootsubo},\altaffilmark{1,2}\altemailmark\orcid{0000-0002-5413-3680} \email{ootsubo23744@med.uoeh-u.ac.jp}
 Hideyo \textsc{Kawakita},\altaffilmark{3}\orcid{0000-0003-2011-9159}
 Tadafumi \textsc{Takata},\altaffilmark{4,5}\orcid{0000-0002-6592-4250}
 Junko \textsc{Furusawa},\altaffilmark{4}\orcid{0000-0002-1968-5762}
 Hisanori \textsc{Furusawa},\altaffilmark{4,5}\orcid{0000-0002-6174-8165}
 Tsuyoshi \textsc{Terai},\altaffilmark{6}\orcid{0000-0003-4143-4246}
 Toshihiro \textsc{Kasuga},\altaffilmark{4,8}\orcid{0000-0001-5903-7391}
 Akira \textsc{Keida},\altaffilmark{4}\orcid{0009-0005-4191-4540}
 Yoshiharu \textsc{Shinnaka},\altaffilmark{3}\orcid{0000-0003-4490-9307}
 Fumi \textsc{Yoshida},\altaffilmark{1,2}\orcid{0000-0002-3286-911X}
 and
 Seitaro \textsc{Urakawa}\altaffilmark{7}\orcid{0000-0001-7501-8983}
}
\altaffiltext{1}{University of Occupational and Environmental Health, Japan, 1-1 Iseigaoka, Yahata, Kitakyusyu, Fukuoka 807-8555, Japan}
\altaffiltext{2}{Planetary Exploration Research Center, Chiba Institute of Technology, 2-17-1 Tsudanuma, Narashino, Chiba 275-0016, Japan}
\altaffiltext{3}{Koyama Space Science Institute, Kyoto Sangyo University, Motoyama, Kamigamo, Kita-ku, Kyoto  603-8555, Japan}
\altaffiltext{4}{National Astronomical Observatory of Japan, National Institutes of Natural Sciences(NINS), 2-21-1 Osawa, Mitaka, Tokyo 181-8588, Japan}
\altaffiltext{5}{Astronomical Science Program, The Graduate University for Advanced Studies, SOKENDAI, 2-21-1 Osawa, Mitaka, Tokyo 181-8588, Japan}
\altaffiltext{6}{Subaru Telescope, National Astronomical Observatory of Japan, 650 North A`ohoku Place, Hilo, HI 96720, USA}
\altaffiltext{7}{Bisei Spaceguard Center, Japan Spaceguard Association, 1716-3 Okura, Bisei-cho, Ibara, Okayama 714-1411}
\altaffiltext{8}{Nishi-Harima Astronomical Observatory, Center for Astronomy, University of Hyogo, 407-2, Nishigaichi, Sayo, Hyogo, 679-5313, Japan}





\maketitle

\begin{abstract}
We present an observational study of the nucleus of comet 28P/Neujmin at a heliocentric distance exceeding 10~au, where coma contamination is effectively minimized.  Observations were conducted in the $g$, $r$, and $y$ bands with the Hyper Suprime-Cam (HSC) on the 8.2-m Subaru Telescope. The measured colors, $g - r = 0.67\pm0.17$ and $r - y = 0.41\pm0.19$, yield a spectral index of $S' = 8.8\pm4.2\%/100$~nm, comparable to that of D-type asteroids. 
By incorporating new observational data at a phase angle $\alpha = \timeform{0.334D}$ with previous observations, we determined the phase function for the nucleus of 28P and confirmed an opposition surge at small phase angles. The derived opposition effect amplitude depends on the adopted phase coefficient, which is uncertain due to potential systematic effects in multi-apparition phase curves. Nevertheless, even under conservative assumptions, the opposition effect of 28P suggests a larger coherent backscattering contribution than is typical for C- and D-type asteroids. The Subaru HSC observations suggest that, although the nucleus color resembles that of D-type asteroids, the surface microstructure of comet 28P's nucleus likely differs from those of C- and D-type asteroids. Future single-apparition observations covering a wide phase angle range from near-opposition to larger angles, combined with polarimetric measurements, will be essential to definitively establish the physical mechanisms responsible for the opposition effects of cometary nuclei.
\end{abstract}


\section{Introduction}

Traditional models of the Solar System formation postulated that ``icy'' comets and ``rocky'' asteroids possess substantially different physical properties and distinct origins. 
However, recent observations have revealed that the boundary between comets and asteroids is ambiguous. 
Some asteroids contain water and hydrated minerals \citep{Rivkin2002,Usui2019,Kitazato2019,Hamilton2019}, while signs of hydrated silicates have been detected on comet nuclei \citep{Ootsubo2021}. 
Understanding which asteroid types share properties with cometary nuclei can provide valuable insights into their origin and evolution.

The opposition effect (OE) in a magnitude-phase curve serves as a key indicator for determining which taxonomic group of asteroids, if any, cometary nuclei most closely resemble. 
The phase curve, which represents the dependence of brightness on the phase angle, provides insights into the photometric and structural properties for the surfaces of comets and asteroids. 
The OE results from one or both of two mechanisms 
\citep[and references therein]{Hapke2012}. 
The shadow-hiding opposition effect (SHOE) is associated with shadows cast by regolith particles that disappear as the phase angle approaches \timeform{0D}. 
The coherent backscattering opposition effect (CBOE) arises from the constructive interference of multiple scattered light rays. 

These two mechanisms can be distinguished by their characteristic signatures in the phase curve:
the CBOE produces sharp and narrow peaks ($<$1${\degree}$--2${\degree}$ wide) at very small phase angles, while the SHOE results in broader features several degrees wide or more \citep{Belskaya2000,Hapke2012}.
The amplitude and angular width of OE can help identify which mechanism dominates the reflectance of target objects.

For asteroids, \citet{Belskaya2000} studied the opposition effect for 33 asteroids comprising various spectral types.
Certain asteroid types (e.g., S- and M-types) demonstrate a rapid increase in brightness near opposition, referred to as the opposition surge, and exhibit significant CBOE contributions (e.g., \cite{Belskaya2000, Hasegawa2014}). 
On the other hand, dark asteroids (e.g., C-, F-, and D-types) typically show little or no opposition effect.

To date, the only detailed study of the OE on cometary nuclei comes from the {\it in situ} observations of 67P/Churyumov-Gerashimenko (hereafter 67P) by the Rosetta spacecraft \citep{Ciarniello2015, Hasselmann2017, Masoumzadeh2017, Masoumzadeh2019}, although the OE has been observed in cometary coma dust (cf. \cite{Rosenbush2002}). These Rosetta studies found that the opposition effect of 67P is primarily controlled by SHOE.
Since comet nuclei typically have low albedo ($\sim$0.04), similar to that of C- and D-type asteroids, one might expect similar opposition effects. Comparing the OEs of comet nuclei and asteroids can offer a key approach to understanding whether comet nuclei share the same surface properties as these dark primitive asteroids, or differ in ways that reflect their distinct evolutionary histories.

However, information on the surface properties of comet nuclei remains ambiguous.
When a comet approaches the Sun and becomes active, a coma develops and conceals the comet nucleus deep within. 
Thus, no ground-based observations have definitively confirmed the presence of the OE in comet nuclei \citep{Donaldson2023}. 
To observe comet nuclei through ground-based observations, rather than relying solely on rare spacecraft missions, comets must be observed when they are at a significant distance ($>$5~au) from the Sun and/or when their activity is extremely low.

To address this observational challenge, observations of comet nuclei at large heliocentric distances using large-aperture telescopes are essential.
The Hyper Suprime-Cam (HSC), a prime focus camera for the 8.2-m Subaru Telescope, is ideal for this research objective \citep{Miyazaki2018, Komiyama2018}.
In our investigation, we found a Jupiter-family comet 28P/Neujmin (hereafter, 28P) in the HSC data, which was observed near aphelion and near opposition (phase angle $\alpha < 1{\degree}$) at a heliocentric distance exceeding 10~au from the Sun, where the sublimation-driven activity of not only water but also carbon dioxide becomes negligible. This represents a rare opportunity to directly observe a cometary nucleus and investigate its surface properties from the ground. 

Comet 28P/Neujmin is a short-period comet with an orbital period of 18.4~years, an eccentricity of 0.774, an inclination of \timeform{14.3D}, and an aphelion distance of 12.38~au. The Tisserand invariant, $T_{J}$, is 2.168, classifying 28P as a Jupiter-family comet (values are taken from NASA JPL/Small-Body Database Lookup\footnote{\url{https://ssd.jpl.nasa.gov/tools/sbdb_lookup.html#/?sstr=28P}}).
The nucleus has a rotational period of $12.75\pm0.03$~hrs, a peak-to-peak amplitude of 0.45~mag, suggesting an elongated shape with axis ratio $\geq$1.5, and an estimated effective radius of $r_{\rm N} = 10.7$~km with albedo $p_{\rm V} = 0.026$ \citep{Campins1987, Delahodde2001, Lamy2004}.

Previous observations of comet 28P in the $R$ band suggested that the opposition surge began to appear at phase angles around \timeform{1.5D} \citep{Delahodde2001}. However, their dataset lacked reliable $R$-band data points for phase angles smaller than \timeform{0.8D}, preventing a definitive conclusion. 
The CBOE peak can easily be distinguished from the rest of the phase curve by its narrow width ($<$1${\degree}$--2${\degree}$).
Observations at small phase angles near opposition (less than 1${\degree}$) are essential to constrain the properties of the OE on comet nuclei.
In this paper, we present a more detailed quantitative study of the OE for 28P based on the Subaru HSC observations at $\alpha\sim\timeform{0.33D}$ in the $r$ band and $\sim$\timeform{0.52D} in the $g$ band, which are closer to opposition than those of \citet{Delahodde2001}. 

The goal of this paper is to study physical properties of the surface of comet 28P, comparing them with those of asteroids and comet 67P, focusing specifically on the OE and color. The 28P data extracted from the Subaru HSC are described in Section 2. In Section 3, we analyze the HSC data and derive the phase curve of comet 28P. The results of the OE of 28P derived from its phase curve and the comparison of OE between 28P and asteroids or comet 67P are discussed in Section 4.

\section{Data}\label{sec:2}

For this study, we utilized archival observations of comet 28P obtained with the HSC on the 8.2-m Subaru Telescope. 
The HSC has a 1.5-degree-diameter field of view and a pixel scale of \timeform{0.168''} \citep{Miyazaki2018, Komiyama2018}, enabling high-quality photometric data acquisition even for faint and/or distant small Solar System bodies (SSSBs).

We utilized archival data from the HSC Subaru Strategic Program (HSC-SSP; \cite{Aihara2018}),
specifically from Public Data Release 3 (PDR3),\footnote{\url{https://hsc-release.mtk.nao.ac.jp/doc/index.php/available-data__pdr3/}} published in August 2021 \citep{AiharaPDR3}.
The survey covers 1400 deg$^2$ of the sky with the full depth of $\sim$26~mag at 5$\sigma$ level depending on the filter \citep{AiharaPDR3}.
To study physical properties of 28P, such as colors and phase curve parameters, we extracted and identified known SSSBs using CCD image data and catalogs of detected sources in the HSC filter bands ($g$, $r$, $i$, $z$, and $y$).
Exposure times for individual frames were 150--300~s in the $g$ and $r$ bands and 200--300~s in the $i$, $z$, and $y$ bands. 
The released data were processed using hscPipe v8, a data reduction and analysis pipeline developed by the HSC collaboration team \citep{Bosch2018,Bosch2019,AiharaPDR3}. This pipeline produces primary data products, including calibrated CCD exposure images (CORR) and source catalogs (SRC) that contain various measured quantities, such as fluxes derived using multiple photometry algorithms and quality flags for detected sources in each CCD. Further details on the HSC-SSP PDR3 can be found in \citet{AiharaPDR3}.

To identify observations of comet 28P, we searched the HSC-SSP PDR3 data using the comet's predicted ephemeris coordinates at the time of each observation and verified the presence of corresponding detected sources in the SRC catalog. Daily ephemerides for 28P were retrieved from the NASA JPL/Horizons system.\footnote{\url{https://ssd.jpl.nasa.gov/horizons/}} 
A brief explanation of the search methodology is available in \citet{Ootsubo2025}. 

We identified observational data of 28P for three nights in the PDR3 data: February 12, March 7, and March 9, 2016. On February 12, HSC observed 28P with eight exposures of 200~s in the $y$ band. The heliocentric ($r_h$) and geocentric ($\Delta$) distances of the comet on February 12 were 10.574~au and 9.718~au, respectively, after its aphelion passage. The phase angle $\alpha$, the Sun-comet-observer angle, is $\timeform{2.779D}$--$\timeform{2.784D}$. On March 7, the comet was observed with three exposures of 150~s in the $g$ band at $r_h = 10.515$~au, $\Delta = 9.527$~au, and $\alpha = \timeform{0.518D}$--$\timeform{0.524D}$. On March 9, it was observed with five exposures of 150~s in the $r$ band at $r_h$ = 10.510~au, $\Delta = 9.519$~au, and $\alpha = \timeform{0.326D}$--$\timeform{0.339D}$. The daily average phase angles $\alpha_\mathrm{avg}$ for observations in the $y$, $g$, and $r$ bands are $\timeform{2.781D}$, $\timeform{0.521D}$, and $\timeform{0.334D}$, respectively. The observation conditions for comet 28P in the PDR3 dataset are summarized in Table \ref{tab:1stlist}.

\begin{figure}
 \begin{center}
  \includegraphics[width=8cm]{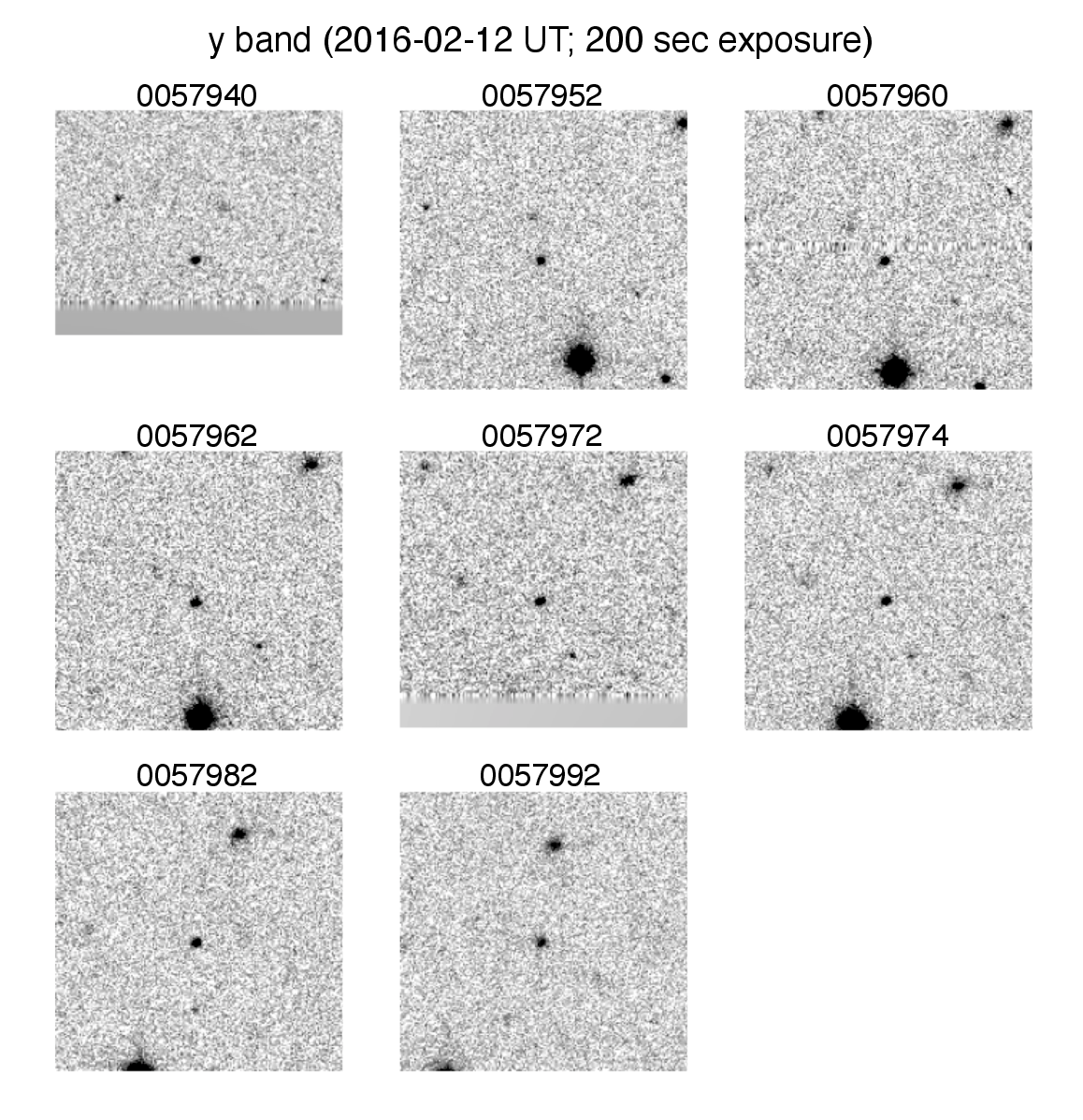} 
 \end{center}
 \begin{center}
  \includegraphics[width=8cm]{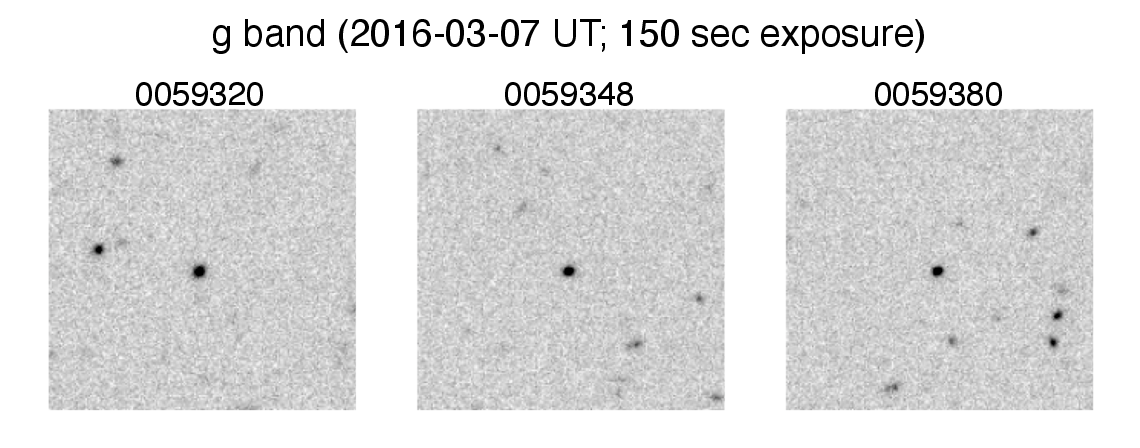} 
 \end{center}
  \begin{center}
  \includegraphics[width=8cm]{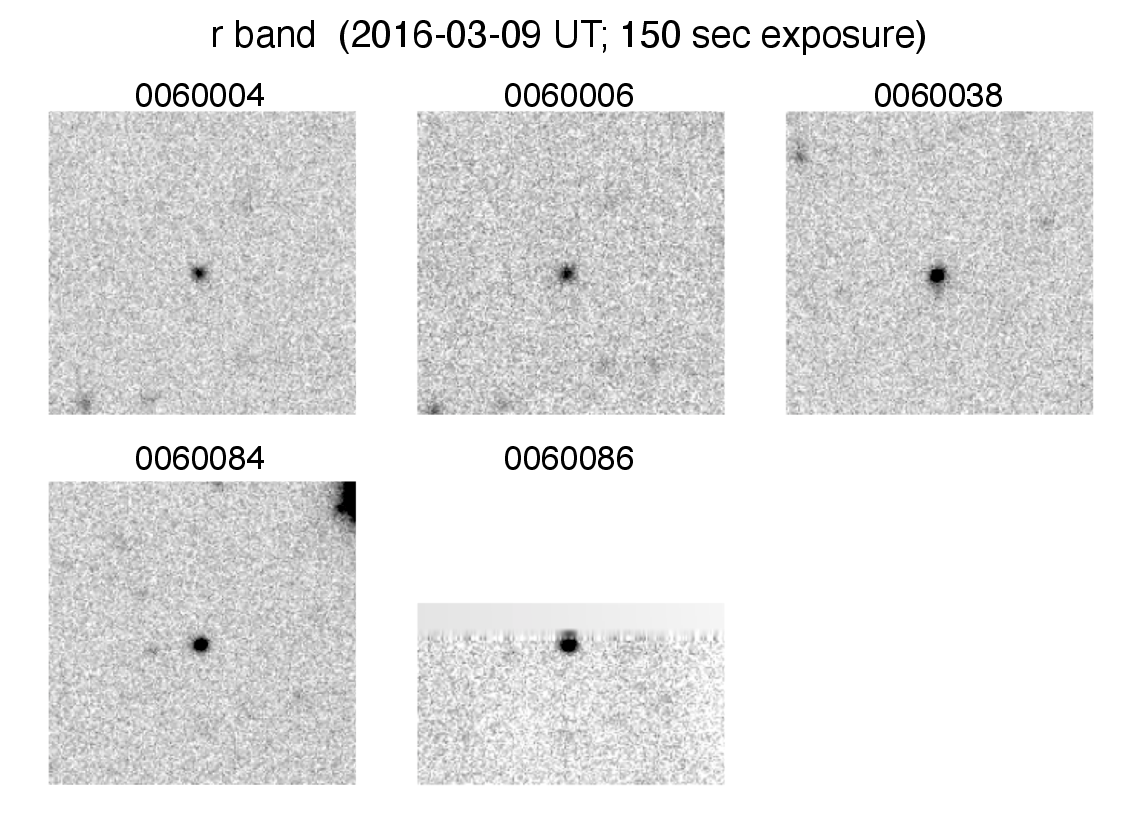} 
 \end{center}
\caption{Images of a $\timeform{30"}\times\timeform{30"}$ region surrounding comet 28P detected in the HSC-SSP data: Eight $y$-band images (top), three $g$-band images (middle), and five $r$-band images (bottom). Each image is labeled with the visit ID (7 digits) of the HSC observation. All images are displayed with North up and East to the left.
{Alt text: Images of comet 28P with no evidence of a resolved coma.}}
\label{fig:images}
\end{figure}

The quality of the comet images was visually inspected using the CORR exposure images. Figure \ref{fig:images} presents the CORR images of a $\timeform{30"}\times\timeform{30"}$ region centered on comet 28P. All data for 28P were obtained during photometric nights. The comet appears near the edge of the CCD for visit 57940 ($y$ band) and visit 60086 ($r$ band), and near an anomaly pattern for visit 57960 ($y$ band). These three frames were excluded from the following analysis. In all other images, comet 28P appeared point-like, with no evidence of a resolved coma. 

\begin{figure*}[t]
 \begin{center}
  \includegraphics[width=17cm]{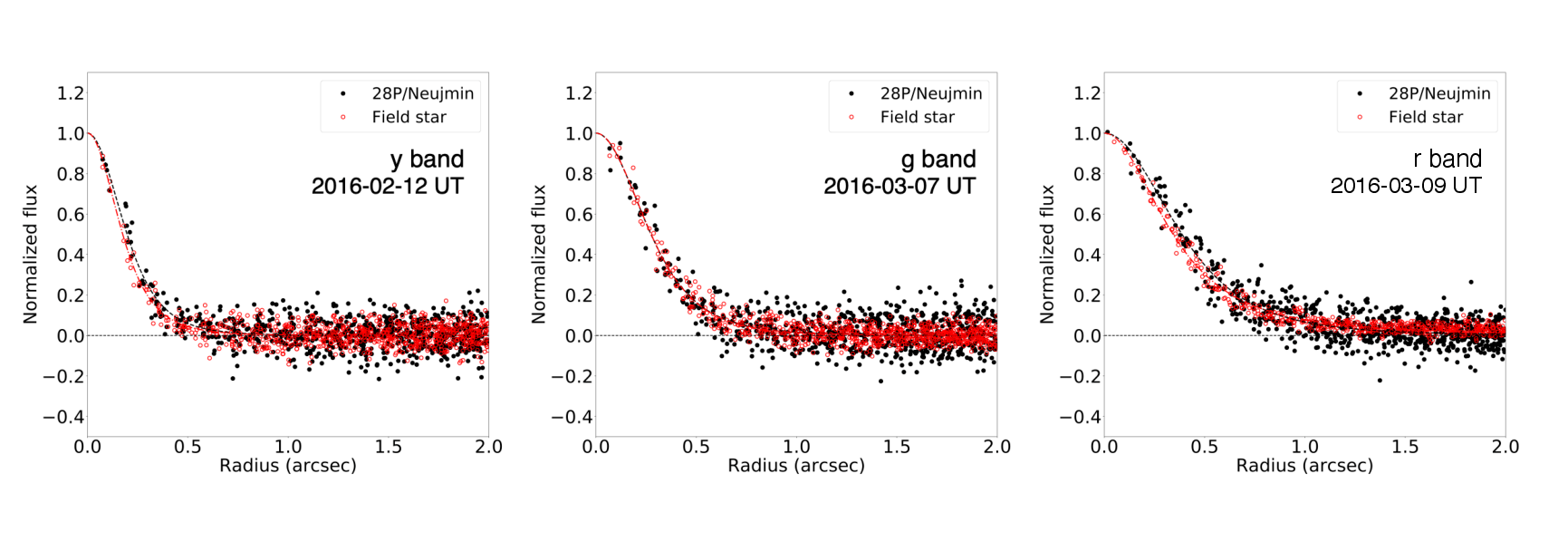} 
 \end{center}
\caption{Radial profiles of comet 28P (black solid circles) compared to those of field stars (red open circles) in the $y$ (visit 57952), $g$ (visit 59320), and $r$ (visit 60038) bands, from left to right. The radial profiles of 28P (black dashed lines) are in good agreement with the point spread function (red dash-dot lines), and no apparent coma structure is detected.
{Alt text: Three plots showing radial profiles of the comet and nearby field stars. In all panels, the x-axis represents the radius from the center of the object, ranging from 0 to 2 arcsec. The y-axis shows the normalized flux, with values normalized to the peak.}}
\label{fig:profile}
\end{figure*}

Although 28P is a moving object and the HSC-SSP observations did not employ non-sidereal tracking, the comet was expected to appear point-like in the HSC images as it was observed at distances beyond 10~au from the Sun and its motion was slow. The comet's apparent motion during the observations was \timeform{0.55''}--\timeform{0.60''} over 150 or 200-s exposures. The seeing during the observations of 28P ranged from \timeform{0.6''} to \timeform{1.2''} in full width at half maximum (Table \ref{tab:1stlist}). Consequently, trailing was barely detectable during the exposures, and all images of the comet appeared as point-like sources. The radial profile of 28P is compared with that of a nearby field star detected in the same CCD. Figure \ref{fig:profile} presents the typical radial profiles of 28P and the field stars in the $y$ (visit 57952), $g$ (visit 59320), and $r$ (visit 60038) bands. The similarity in the radial profiles of 28P and the field stars confirms the absence of observable activity around the nucleus of 28P.

Finally, we utilized the data from six $y$-band, three $g$-band, and four $r$-band observations for the analysis and discussion (Table \ref{tab:1stlist}). For the photometry of 28P, we used the magnitudes derived from the fluxes in the SRC catalog measured with a 12-pixel (i.e., $\sim$\timeform{2.0''}) radius aperture ($m_{x}$; $x$ represents filter bands: $g$, $r$, and $y$), which is the same configuration employed for estimating the zero-point magnitudes. It should be noted that the magnitudes of objects in the SRC catalog are expressed in the AB magnitude system (Table \ref{tab:1stlist}). 

\section{Analysis and Results}\label{sec:3}

\subsection{Mean magnitude and color of 28P}

We obtained the apparent magnitude of comet 28P for each observation from the SRC catalog. Then, we computed the mean magnitudes of the comet to analyze its color and phase curve.
The reduced magnitude, $m_{x}(1,1,\alpha)$, and absolute magnitude, $H_{x}$,  in $x$ band can be calculated using the following equations:
\begin{equation}
m_{x}(1,1,\alpha) = M_{x} - 5\log(r_h\Delta),
\end{equation}
\begin{equation}
H_{x} = m_{x}(1,1,0) = m_{x}(1,1,\alpha) - \alpha\beta,
\end{equation}
where $x$ is the filter band ($g$, $r$, or $y$), $M_{x}$ is the mean magnitude of the comet in $x$ band, $r_h$ is the heliocentric distance (in au), $\Delta$ is the geocentric distance (in au), $\alpha$ is the phase angle (the Sun-comet-observer angle in degrees), and $\beta$ is the phase coefficient (in mag\,degree$^{-1}$). 
We also derive the reduced magnitude at the phase angle $\alpha = \timeform{0.334D}$ (the average phase angle of the four $r$-band observations) for each band to compare with the previous results at $\alpha = \timeform{0.3D}$ \citep{Belskaya2000}:
\begin{equation}
m_{x}(1,1,0.334) = M_{x} - 5\log(r_h\Delta) - (\alpha - 0.334)\beta.
\end{equation}

For the phase coefficient, $\beta$, of comet 28P, previous studies have suggested values in the range of 0.02--0.05~mag\,degree$^{-1}$ (e.g., $0.034\pm0.012$ in \cite{Jewitt1988}, $0.025\pm0.006$ in \cite{Delahodde2001}, and $\sim$0.05~mag\,degree$^{-1}$ in \cite{Schleicher2022}). Typically, $\beta$, which is used for comet observations, is derived at large phase angles using a linear fit without an OE. For consistency in our analysis and discussion of comet 28P, which shows the OE, we adopt 0.022 for the $\beta$ of comet 28P derived at large phase angles with a function considering the OE, as will be derived later as $b$ in equation (\ref{Shevchenko}) in Section 3.3.1 of this paper. The HSC data used in this study were obtained at small phase angles (\timeform{0.33D}--\timeform{2.78D}), and the variation in $\beta$ within the range of 0.02--0.05 has a negligible effect on the results and the following discussion.

In principle, the HSC-SSP observations were conducted with one or a few filters per night. For the nights when 28P was detected, observations were carried out using a single filter each night. Brightness variations due to the rotation of the comet nucleus were detected each night of the observation. Previous studies suggest that comet 28P exhibits a rotational lightcurve with an asymmetric, double-peaked shape, a rotational period of $\sim$12.75 hrs, and a peak-to-peak amplitude of $\sim$0.45~mag, indicating that the elongation of the nucleus may be $>$1.5 \citep{Delahodde2001}. 
The duration of the HSC observation for 28P each night (1.5--3.5~hrs) was less than half of the comet's rotational phase, likely preventing the capture of both maximum and minimum brightness. However, the observed magnitude variations ($\Delta m_x$) were approximately 0.23, 0.40, and 0.34~mag in the $g$, $r$, and $y$ band, respectively (Table \ref{tab:1stlist}). This suggests that the average of the observed magnitudes is unlikely to differ significantly from the mean magnitude, $M_{x}$, which is expected to be near the middle of the 0.45~mag range.
We adopt an average of the observed data as the mean magnitude of 28P in each band. 

Since the HSC data for the comet in the $g$, $r$, and $y$ bands were likely obtained at different rotational phases, lightcurve variation of the nucleus could introduce additional uncertainty in determining the color and the spectral index. 
To evaluate the impact of these variations on our color measurements, we performed a Monte Carlo simulation using 1,000,000 synthetic lightcurves with randomized initial rotational phases \citep{Terai2018,Sakugawa2018}. These lightcurves were generated based on a double-peaked sinusoidal brightness fluctuation with a rotational period of $12.75\pm0.03$~hrs and an amplitude of $0.45\pm0.05$~mag \citep{Delahodde2001}. 
The uncertainty in a color due to rotation ($\sigma_\mathrm{rot}$) was quantified as the standard deviation of the color offsets measured with the same cadence as the actual observations. Our simulation indicates that $\sigma_\mathrm{rot}$ is 0.148~mag for the $g - r$ color and 0.154~mag for the $r - y$ color. In brief, the typical $\sigma_\mathrm{rot}$ value corresponds to 0.15~mag, which should be incorporated into the error estimation of the colors.

The mean magnitudes, $M_{x}$, the absolute magnitudes, $H_{x}$, and the reduced magnitudes at the phase angle of $\alpha = \timeform{0.334D}$, $m_{x}(1,1,0.334)$, are presented in Table \ref{tab:mags}. The colors $g - r$ and $r - y$, derived from $m_{g}(1,1,0.334)$, $m_{r}(1,1,0.334)$, and $m_{y}(1,1,0.334)$, are $0.67\pm0.17$ and $0.41\pm0.19$, respectively. In the derivation of the mean magnitudes and their uncertainties, we considered the possibility that the $g$-, $r$-, and $y$-band observations sampled different rotational phases of the cometary nucleus. In the following sections, we analyze and discuss the properties of the comet nucleus based on the magnitudes and uncertainties obtained above.

\subsection{Reflectance spectrum and albedo}

We derive the reflectance of the comet nucleus using the following relation between the observed mean magnitude, radius, and albedo of the comet nucleus \citep{Meech2004,Lamy2011}:
\begin{equation}
p_{x}r_\mathrm{N}^2 = 2.238 \times 10^{22}~\Delta^2~r_h^2~10^{0.4(m_{\Sol,x}-M_{x}+\alpha\beta)},
\end{equation}
where $p_x$ is the geometric albedo in the $x$ band, $r_\mathrm{N}$ is the radius of the comet nucleus (in meters), and $m_{\Sol,x}$ is the solar magnitude in the $x$ band.
Solar magnitudes in the AB system for the HSC filter bands in the $g$, $r$, and $y$ bands are taken from \citet{Willmer2018}\footnote{\url{http://mips.as.arizona.edu/~cnaw/sun.html}}:
$m_{\Sol, g} = -26.51$,
$m_{\Sol, r} = -26.93$, and
$m_{\Sol, y} = -27.07$.
Assuming $r_\mathrm{N} = 10.7$~km \citep{Lamy2004} and $\beta = 0.022$, we obtained
$p_g = 0.037\pm0.005$,
$p_r = 0.046\pm0.003$, and 
$p_y = 0.059\pm0.007$.
Although wavelength differences should be taken into consideration, 
the derived albedo value ($p_g =0.037\pm0.005$) shows marginal agreement with 
previously reported visible geometric albedo ($p_{\rm V} = 0.026$; \cite{Lamy2004}).

The spectral index of 28P, derived from the $g$, $r$, and $y$-band data, is
$S' = 8.8\pm4.2$\%/100~nm, aligning with previous results around the $V$ and $R$ bands:
$S' = 9.1\pm1.9$\%/100~nm \citep{Delahodde2001},
$13\pm4$\%/100~nm \citep{Jewitt1988}, 
$10.3\pm3.0$\%/100~nm, and $13.3\pm2.0$\%/100~nm \citep{Campins2007}.

\subsection{Phase curve and opposition effect}

\citet{Delahodde2001} discussed the phase curve of 28P based on the reduced magnitude using the Vega magnitude of the Bessel $R$-band, $m_{R}(1,1,\alpha)$,
for phase angles ranging from $\sim$\timeform{0D} to \timeform{15D}. Since the HSC data were obtained using a different filter system, we convert the observed HSC magnitudes to the Bessel $R$ magnitude as outlined below. 
To perform this conversion from HSC $g$- and $r$-band magnitudes to Bessel $R$, 
we use the same procedure as in Section 6.2 of \citet{Furusawa2018}, using a stellar spectral library from \citet{GS1983}. 
In this case, we convolve the spectra with the HSC $g$ and HSC $r$ and the Bessel $R$ bandpasses to derive the color terms for the conversion. 

The HSC $g$- and $r$-band magnitudes of 28P at the same phase angle are required to derive the $m_{R}(1,1,\alpha)$. 
First, we estimate the $g$-band magnitude at $\alpha = \timeform{0.334D}$ (the average phase angle of the $r$-band observations), $m_{g}(1,1,0.334)$, from the observed HSC magnitude at $\alpha = \timeform{0.521D}$ (the average phase angle of the $g$-band observations). 
Assuming a phase coefficient $\beta$ of 0.022 and taking into consideration the uncertainty $\sigma_\mathrm{rot}$,
the magnitudes and colors of 28P are as follows:
$m_{r}(1,1,0.334) = 12.15\pm0.08$, 
$m_{g}(1,1,0.334) = 12.82\pm0.06$, and
$(g - r)_\mathrm{HSC} = 0.67\pm0.17$.

To obtain the $V$ and $R$ magnitudes in AB system ($m_{V}(1,1,\alpha)_\mathrm{AB}$ and $m_{R}(1,1,\alpha)_\mathrm{AB}$, respectively) from the HSC $g$- and $r$-band magnitudes, we adopt conversion equations (conversion equations for $B$, $V$, $R$, and $I$ bands are shown in Appendix):

\begin{equation}
  \begin{split}
    m_{V}(1,1,0.334)_\mathrm{AB} &= m_\mathrm{g}(1,1,0.334) - 0.0140 \\
    &\quad - 0.5689 (g - r)_\mathrm{HSC} \\ 
    &\quad - 0.0050 ((g - r)_\mathrm{HSC})^{2}
  \end{split}
\end{equation}

\begin{equation}
  \begin{split}
	m_{R}(1,1,0.334)_\mathrm{AB} &= m_\mathrm{r}(1,1,0.334) + 0.0192 \\
	&\quad - 0.1346 (g - r)_\mathrm{HSC} \\
	&\quad + 0.0113 ((g - r)_\mathrm{HSC})^{2}
  \end{split}
\end{equation}

Finally, to derive the Vega magnitude, 0.011 and 0.199 are subtracted from $m_{V}(1,1,0.334)_\mathrm{AB}$ and $m_{R}(1,1,0.334)_\mathrm{AB}$, respectively (\cite{Fukugita1995}; see Appendix).
Thus, we obtain the magnitudes of 28P at $\alpha = \timeform{0.334D}$: $m_{R}(1,1,0.334) = 11.84\pm0.19$ and $m_{V}(1,1,0.334) = 12.43\pm0.15$.
The resultant color of 28P is $V - R = 0.59\pm0.24$. This result is redder than $V - R = 0.45\pm0.05$ reported by \citet{Delahodde2001}, but remains consistent within the error margins.


\begin{figure}
 \begin{center}
  \includegraphics[width=8cm]{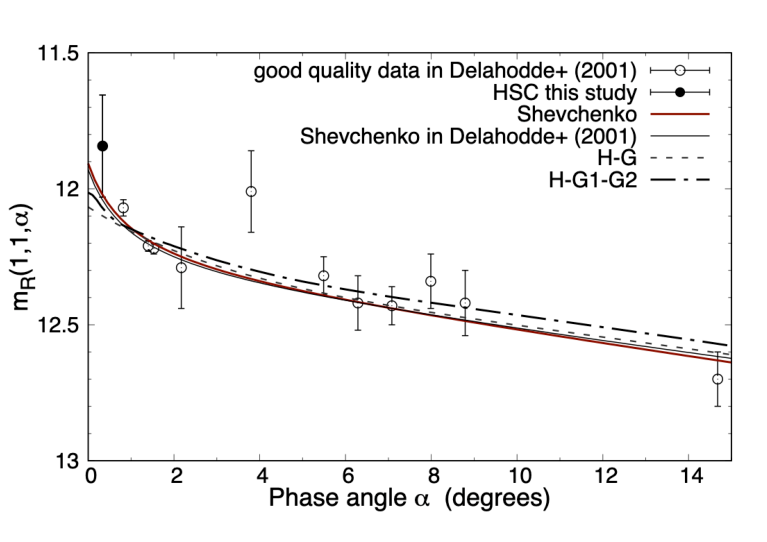} 
 \end{center}
 \caption{Phase curve of comet 28P in the R band showing the HSC data (filled circle) and previous observations (open circles; \cite{Delahodde2001}) together with fitted phase function models: Shevchenko model for the HSC data with previous data (red solid line), Shevchenko model in \cite{Delahodde2001} (black solid line), IAU H-G model (dashed line), and IAU H-G$_1$-G$_2$ model (dash-dotted line). 
 {Alt text: Comparison of fitted models with observed data. The x-axis represents phase angles ranging from 0 to 15 degrees. The y-axis shows reduced magnitudes in the $R$-band.}}
\label{fig:phasecurve}
\end{figure}

\subsubsection{Comparison with asteroids}

Figure \ref{fig:phasecurve} shows the phase curve of comet 28P in the $R$ band. The derived HSC magnitude $m_{R}(1,1,0.334)$ is plotted together with the $m_{R}(1,1,\alpha)$ from \citet{Delahodde2001}. We examine the phase curve characterization in the $R$-band for comet 28P. \citet{Delahodde2001} discussed the phase curve and OE for asteroids using the IAU $H$-$G$ model \citep{Bowell1989} and the Shevchenko model \citep{Shevchenko1996}. We also interpret the OE using the IAU-adopted $H$-$G_1$-$G_2$ system \citep{Muinonen2010}, which includes the term for the OE, in addition to the Shevchenko model and the IAU $H$-$G$ model. \\

\begin{itemize}
  \item Shevchenko model
\end{itemize}
The Shevchenko phase function model is a simple three-parameter empirical function that was proposed by \citet{Shevchenko1996}.
This function has the form:
\begin{equation}\label{Shevchenko}
m_{R}(1,1,\alpha) = m_{R}(1,1,0)-a/(1+\alpha)+b\,\alpha,
\end{equation}

\noindent where $a$ is a parameter characterizing the OE amplitude and $b$ is the parameter describing the linear part of the magnitude phase dependence.
In \citet{Delahodde2001}, the phase curve of 28P was fitted using the Shevchenko model with the parameters, 
$m_{R}(1,1,0) = 12.35\pm0.03$, $b = 0.020\pm0.008$, and $a = 0.42\pm0.05$.
In the present study,  we perform a model fit on the phase curve of the data, incorporating the additional HSC results (figure \ref{fig:phasecurve}). 
The fit yields the following parameters: 
$m_{R}(1,1,0) = 12.34\pm0.05$,  $b = 0.022\pm0.012$, and $a = 0.43\pm0.13$.
There is no significant change in the results.\\

\begin{itemize}
  \item IAU H-G system
\end{itemize}
The IAU-adopted $H$-$G$ model is a simple two-parameter empirical function \citep{Bowell1989}.
This is the most common phase function for asteroids and has the form:
\begin{equation}
  \begin{split}
	m_{R}(1,1,\alpha) &= H - 2.5\log(1-G)\exp[-3.33\tan^{0.63}(\alpha/2)] \\
                  &\quad + G\exp[-1.87\tan^{1.22}(\alpha/2)],
  \end{split}
\end{equation}

\noindent where $H$ is the absolute magnitude and $G$ is the so-called slope parameter,
which describe the shape of the magnitude phase function.
In \citet{Delahodde2001}, the phase curve of 28P was fitted using IAU $H$-$G$ model, with the parameters 
$H = 12.07\pm0.03$ and $G = 0.41\pm0.08$. 
Our result yield $H = 12.07\pm0.03$ and $G = 0.45\pm0.12$ (figure \ref{fig:phasecurve}). 
There is no change in $H$ and $G$ is found to be only slightly larger.\\

\begin{itemize}
\item IAU H-G$_1$-G$_2$ system
\end{itemize}
Next, we adopt the three-parameter IAU $H$-$G_1$-$G_2$ system, which includes the OE term \citep{Muinonen2010}. 
In this model, the parameter $H$ represents the absolute magnitude, with $G_1$ and $G_2$ are the inclination parameters of the phase curve \citep{Penttila2016,Ieva2022}.
The phase curve in this model is described by the equation: 
\begin{equation}
  \begin{split}
	m_{R}(1,1,\alpha) &= H - 2.5\log[G_1\Phi_1(\alpha) + G_2\Phi_2(\alpha)\\
	              &\quad + (1 - G_1 - G_2) \Phi_3(\alpha)],
  \end{split}
\end{equation}

\noindent where $\Phi_1$ and $\Phi_2$ are the phase functions associated with the linear part of the curve, and $\Phi_3$ is associated with the OE. 

The model fitting tool based on  \citet{Penttila2016} is available on the web.\footnote{\url{https://psr.it.helsinki.fi/HG1G2/}} Using this tool, we fit the phase curve of 28P with the $H$-$G_1$-$G_2$ model. The result yeilds $H = 11.99$, $G_1 = 0.0668$ and $G_2 = 0.623$ (figure \ref{fig:phasecurve}). Compared to asteroids, the surface properties of the 28P nucleus are closer to those of E-type asteroids than D- or C-type.

\subsubsection{Comparison with comet 67P}

Observed data obtained with spacecraft are often expressed in terms of the radiance factor, denoted as $I/F$, rather than magnitudes (e.g., \cite{Masoumzadeh2017, Masoumzadeh2019}). The radiance factor is defined as the ratio of the radiance of an illuminated surface to the radiance from a normally illuminated Lambert surface at the same observer-object distance. To reproduce and interpret the phase curves, empirical models are often used (e.g., \cite{Rosenbush2002}). Following \citet{Rosenbush2002} and \citet{Masoumzadeh2019}, we adopt a four-parameter linear-exponential model to analyze the phase curve of 28P:

\begin{equation}
I/F = I/F_s \exp\left(-\frac{\alpha}{1.45\times\textit{HWHM}}\right) + I/F_b + B\alpha,
\end{equation}

\noindent where $I/F_s$ is the amplitude of OE and is defined as the brightness increase relative to the background brightness $I/F_b$. $B$ is the slope of the linear part, and \textit{HWHM} is the angular width of the OE enhancement.
Following \citet{Rosenbush2002}, we estimated the enhancement factor $\zeta$ as the amplitude of OE,

\begin{equation}
\zeta = \frac{I/F_s + I/F_b}{I/F_b}.
\end{equation}

The observed magnitude is converted to $I/F$ using the following relationship \citep{Bach2019}:

\begin{equation}
-2.5\log_{10}\left(\frac{I}{F}\right) = m_{R}(1,1,\alpha) - R_{\mathrm{mag}\Sol} - 2.5\log_{10}\left(\frac{\pi}{S_\mathrm{proj}}\right) + m_c,
\end{equation}

\noindent where $R_{\mathrm{mag}\Sol} = -27.15$ is the $R$-band Vega magnitude of the Sun at 1~au \citep{Mann2015}, $S_\mathrm{proj}$ is the geometrical cross section in m$^{2}$, and $m_{c} = -5\log_{10}$(1~au/1~m)$ = -55.87$ is a constant to adjust the length unit \citep{Bach2019}. To calculate $S_\mathrm{proj}$, we adopt 10.7~km as the effective radius of 28P \citep{Lamy2004}.

While \citet{Masoumzadeh2017} discussed the Rosetta data of 67P over the phase angle range $\timeform{0D}$--$\timeform{55D}$, the phase angles in our 28P data, including the HSC data and the data from \citet{Delahodde2001}, cover only $\alpha < \timeform{15D}$. \citet{Masoumzadeh2019} analyzed data for small phase angles in the range ($\timeform{0D}$--$\timeform{9D}$), which is closer to our dataset. Thus, we compare our results in the $\timeform{0D}$--$\timeform{9D}$ range with those of \citet{Masoumzadeh2019}. 

\begin{figure}
 \begin{center}
  \includegraphics[width=8cm]{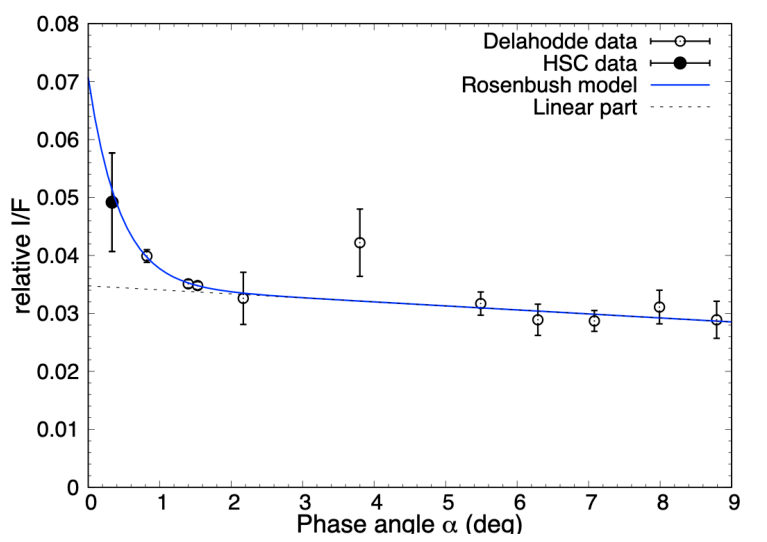} 
 \end{center}
\caption{$I/F$ phase curve of comet 28P in the phase angle of $\timeform{0D}$--$\timeform{9D}$. The blue line represents the fitted linear-exponential model. The dotted lines show a linear component of the model function.
 {Alt text: A line graph showing fitted models plotted against observed data. The x-axis represents phase angles ranging from 0 to 9 degrees. The y-axis shows the relative radiance factor.}}
\label{fig:ioverf}
\end{figure}

Figure \ref{fig:ioverf} shows the $I/F$ phase curve for 28P in the phase angle range of $\timeform{0D}$--$\timeform{9D}$. We fit this phase curve using the linear-exponential model with the Levenberg-Marquardt algorithm. The four parameters, $I/F_{s}$, $I/F_{b}$, \textit{HWHM}, and $B$ were derived, and the best-fit value of $\zeta$ was calculated. 
The resultant values are: 
$I/F_s = 0.036\pm0.023$,
$I/F_b = 0.035\pm0.002$,
$\textit{HWHM} = 0.30\pm0.16$, and
$B = -0.0007\pm0.0003$.
Thus, the OE enhancement facter is $\zeta = 2.04\pm1.33$.

In figure \ref{fig:ioverf}, the best-fit model and the linear component of the function, $I/F_b + B\alpha$, are plotted to clearly illustrate the OE. This model reproduces the OE component more effectively than the IAU models and the Shevchenko model. 

\section{Discussion and conclusion}\label{sec:4}

The Subaru/HSC observations of comet 28P revealed colors of $g - r = 0.67\pm0.17$ 
and $r - y = 0.41\pm0.19$, corresponding to $V - R = 0.59\pm0.24$. 
These colors place 28P within the range typical of D-type asteroids, despite the relatively large error bars. 
The derived spectral index, $S' = 8.8\pm4.2\%/100$~nm, which is also in agreement with previous results obtained around the $V$ and $R$ bands ($S' = 9.1\pm1.9$\%/100~nm; \cite{Delahodde2001})
and lies within the range observed for D-type asteroids ($9.5\pm1.6$\%/100~nm in \cite{Fitzsimmons1994}) and Jupiter Trojans ($8.8\pm1.0$\%/100~nm in \cite{Jewitt1990}).
This slope is notably less steep than the average values of Trans-Neptunian Objects ($\sim$24--28\%/100~nm) \citep{Hainaut2002},
further supporting the classification of 28P as a D-type analog in terms of surface composition.

The OE provides critical insights into the surface structure and scattering properties of comet 28P's 
nucleus. 
By incorporating the new HSC data at $\alpha = \timeform{0.334D}$ with the previous observations
by \citet{Delahodde2001} spanning $\timeform{0.82D} < \alpha < \timeform{14.7D}$,
we have determined a well-constrained phase function for the 28P's nucleus.
Our analysis using the Shevchenko model yields a phase coefficient of $0.022\pm0.012$ for the range $\timeform{5.0D} < \alpha < \timeform{14.7D}$, and confirms the presence of an opposition surge at small phase angles, as suggested by \citet{Delahodde2001}.
Following the methodology of \citet{Belskaya2000}, we derived the OE amplitude relative to the linear extrapolation and the CBOE contribution at $\alpha=\timeform{0.334D}$.
The OE amplitude is $0.33\pm0.03$ and the CBOE contribution is $0.92\pm0.12$. 

A key finding of this study is that comet 28P exhibits OE characteristics that differ markedly
from those of C- and D-type asteroids, despite their similar colors and spectral indices.
\citet{Belskaya2000} derived the OE amplitude and the relative contribution of CBOE at $\alpha=\timeform{0.3D}$ for 33 asteroids.
Comparing our results for 28P with their asteroid sample,
we find that the nucleus of 28P exhibits an OE amplitude comparable to that of S-, M-, and E-type asteroids, and significantly larger than the  mean values for C- and D-type asteroids.
The CBOE contribution of 28P, $0.92\pm0.12$, is also comparable to those observed for higher-albedo S-, M-, and E-type asteroids (typically 0.8--0.9), and is larger than the typical values (0.2--0.6) of low-albedo asteroids.
Using the IAU $H$-$G_1$-$G_2$ model, we obtained the parameters, $H = 11.99$, $G_1 = 0.0668$ and $G_2 = 0.623$. The surface properties of the 28P nucleus are closer to those of E-type asteroids than D- or C-type asteroids in this parameter space \citep{Penttila2016,Ieva2022}. 
These results suggest that the surface microstructure of comet 28P differs significantly from those of 
primitive asteroids, although direct comparisons between comet nuclei and asteroids should be interpreted with caution. 

It should be noted, however, that a critical aspect of this study is the uncertainty in the phase coefficient, 
which directly affects the derived opposition effect amplitude.
Our analysis yields $0.022 \pm 0.012$~mag\,degree$^{-1}$ based on fitting the combined dataset including the multi-apparition observations of \citet{Delahodde2001} obtained between 1985 and 2000, spanning more than two 18.4-year orbital periods. 
In contrast, \citet{Schleicher2022} reported $\sim$0.05~mag\,degree$^{-1}$ from the recent single-apparition observations. 
The lower value reported by \citet{Delahodde2001} may have been affected by brightness variations 
due to the changing projected cross-section of the elongated nucleus across different apparitions and viewing geometries, 
potentially flattening the derived phase slope.

If the intrinsic phase coefficient is indeed steeper (0.05~mag\,degree$^{-1}$), linear extrapolation from $\alpha=\timeform{5.0D}$ (where the opposition effect typically becomes negligible) to $\alpha=\timeform{0.33D}$ would predict a brightness increase of $\sim$0.23~mag ($\sim$0.10~mag for the phase coeffcient of 0.022). This would reduced the OE amplitude and the CBOE contribution to 0.20~mag and $0.62$, respectively, suggesting that the opposition effect of 28P is weaker than initially derived. 
Nevertheless, even with this reduced amplitude, 28P would still exhibit a stronger opposition effect than the mean value for C-type asteroids ($\sim$0.16~mag; \cite{Belskaya2000}), 
placing it among the prominent cases within this low-albedo taxonomic class.

The opposition surge at very small phase angles ($< \sim$1$\degree$) characteristic of CBOE is typically not observed in the phase curves of low-albedo objects \citep{Belskaya2000}. 
Cometary nuclei are generally considered to be low-albedo bodies with characteristics similar to D-type asteroids \citep{Fitzsimmons1994}. 
However, the surface properties of the 28P nucleus differ from those of D-type asteroids in terms of the OE amplitude, CBOE contribution, and $G_1$-$G_2$ relationship. Even if comet 28P and D-type asteroids shared a common origin in the outer Solar System, repeated episodes of cometary activity driven by ice sublimation may have altered the surface microstructure of the comet nucleus, potentially creating conditions more favorable for CBOE than those found on primitive asteroids.

A comparison with in situ observations of comet 67P provides further insight into the surface scattering properties of cometary nuclei. 
We also fitted the phase curve of 28P with the linear-exponential model and derived the amplitude and the width of the OE peak.
The nucleus of 28P shows a larger enhancement factor $\zeta$, $2.04\pm1.35$, and a narrower \textit{HWHM}, $0.30\pm0.16$, compared to 67P. 
The Rosetta results for 67P at 649~nm indicate $\zeta \sim 1.11$--1.31 and $\textit{HWHM} \sim \timeform{1.39D}$--\timeform{3.26D} for the $\alpha = \timeform{0D}$--$\timeform{9D}$ range \citep{Masoumzadeh2019}. 
Although direct comparisons between ground-based observations and spacecraft measurements should be made with caution, these results suggest that the OE of 28P is characterized by a sharper and more pronounced enhancement at very small phase angles. 
Such behavior is more consistent with a dominant contribution from CBOE rather than SHOE, and implies that the surface scattering properties of 28P differ from those inferred for 67P.

The phase coefficient uncertainty is also likely to affect the linear-exponential model parameters.
Our calculations show that if the phase coefficient changes from 0.022 to 0.05~mag~degree$^{-1}$, the linear slope parameter $B$ 
in the Rosenbush model would increases by approximately a factor of 2.5 (from $B = -0.0007$ to $B \sim -0.0017$).
However, the other fitted parameters show much smaller variations: HWHM shows no significant change, 
while $I/F_{\rm s}$ decrease and $I/F_{\rm b}$ increase by only $\sim$0.005 ($\lesssim$15\%), respectively. 
Consequently, the enhancement factor $\zeta$ decreases moderately from 2.04 to $\sim$1.8 by approximately 15\%.
Importantly, even with this reduction, $\zeta$ for 28P remains larger than the values reported for 67P ($\zeta \sim 1.1$--1.3; \cite{Masoumzadeh2019}), indicating that the apparent difference in opposition effect 
characteristics between the two cometary nuclei is robust against the phase coefficient uncertainty.

Despite the phase coefficient uncertainty, our Subaru HSC observations have achieved the smallest phase angle measurements for the nucleus of 28P from ground-based facilities ($\alpha = 0.334\deg$), definitively confirming the presence of an opposition surge. 
The detection of such narrow opposition surge on the dark cometary surface has important implications for understanding surface microstructure, but requires careful observational validation.
The case of asteroid 419 Aurelia serves as a cautionary example: an initially reported narrow opposition surge 
attributed to CBOE and interpreted as evidence for sub-$\mu$m grains \citep{Belskaya2002} was later found to be weaker than originally claimed after improved photometric calibration \citep{Shevchenko2016}. 
Definitive confirmation of CBOE on cometary nuclei requires observations that eliminate the systematic effects associated with multi-apparition datasets and changing viewing geometry.
\citet{Hapke1998} proposed that the angular width of a CBOE peak is proportional to wavelength, but that of a SHOE peak is not, and that a polarization OE should accompany a CBOE. 
Future multi-wavelength photometric observations including phase angles below 1${\degree}$ 
within a single apparition,
combined with polarimetric measurements, will be essential to conclusively confirm the dominance of CBOE and to further constrain the surface properties of cometary nuclei.

\begin{landscape}

\begin{table}
  \tbl{Observing conditions. }{%
  \begin{tabular}{ccccccccccccccc}
      \hline
        DateTime UTC\footnotemark[$a$] & Pointing\footnotemark[$b$] & visit\footnotemark[$c$] & ccd\footnotemark[$d$] & exptime\footnotemark[$e$] & r$_h$\footnotemark[$f$] & $\Delta$\footnotemark[$g$] & filter & airmass & seeing & phase angle\footnotemark[$h$] & FWHM\footnotemark[$i$] & m$_{x}$\footnotemark[$j$] & $m_{x}(1,1,\alpha)$\footnotemark[$k$] \\
           &                  &     &     & [s]     & [au]  & [au]    &        &         & [arcsec] & [degree] & [arcsec] & [mag] & [mag]  \\                
      \hline
      2016-02-12T13:34 & 1503 & 57940 & 49 & 200 & 10.574 & 9.718 & y & 1.10 & 0.60 & -- & -- & -- & -- &  \\
      2016-02-12T13:54 & 1503 & 57952 & 27 & 200 & 10.574 & 9.718 & y & 1.13 & 0.56 & 2.784 & 0.64 & 21.77$\pm$0.10 & 11.71 \\
      2016-02-12T14:10 & 1503 & 57960 & 88 & 200 & 10.574 & 9.718 & y & 1.16 & 0.64 & -- & -- & -- & -- &  \\
      2016-02-12T14:14 & 1503 & 57962 &  0 & 200 & 10.574 & 9.718 & y & 1.17 & 0.60 & 2.783 & 0.63 & 21.69$\pm$0.09 & 11.63 \\
      2016-02-12T14:33 & 1503 & 57972 & 38 & 200 & 10.574 & 9.718 & y & 1.23 & 0.61 & 2.781 & 0.60 & 21.72$\pm$0.10 & 11.66 \\
      2016-02-12T14:37 & 1503 & 57974 & 44 & 200 & 10.574 & 9.718 & y & 1.24 & 0.66 & 2.781 & 0.64 & 22.00$\pm$0.12 & 11.94 \\
      2016-02-12T14:53 & 1503 & 57982 & 84 & 200 & 10.574 & 9.718 & y & 1.29 & 0.72 & 2.780 & 0.76 & 22.03$\pm$0.12 & 11.97 \\
      2016-02-12T15:12 & 1503 & 57992 & 82 & 200 & 10.574 & 9.718 & y & 1.38 & 0.90 & 2.779 & 0.90 & 21.97$\pm$0.11 & 11.91 \\
      2016-03-07T11:25 & 1527 & 59320 & 94 & 150 & 10.515 & 9.527 & g & 1.07 & 0.81 & 0.524 & 0.73 & 22.71$\pm$0.04 & 12.71 \\
      2016-03-07T12:09 & 1527 & 59348 & 54 & 150 & 10.515 & 9.527 & g & 1.12 & 0.61 & 0.521 & 0.58 & 22.82$\pm$0.04 & 12.82 \\
      2016-03-07T12:59 & 1527 & 59380 & 96 & 150 & 10.515 & 9.527 & g & 1.24 & 0.56 & 0.518 & 0.59 & 22.95$\pm$0.04 & 12.94 \\
      2016-03-09T10:03 & 1529 & 60004 & 55 & 150 & 10.510 & 9.519 & r & 1.07 & 1.12 & 0.339 & 0.64 & 22.37$\pm$0.06 & 12.36 \\
      2016-03-09T10:07 & 1529 & 60006 & 61 & 150 & 10.510 & 9.519 & r & 1.07 & 1.25 & 0.339 & 0.64 & 22.29$\pm$0.05 & 12.28 \\
      2016-03-09T12:21 & 1529 & 60038 & 97 & 150 & 10.510 & 9.519 & r & 1.16 & 1.02 & 0.330 & 0.63 & 21.97$\pm$0.03 & 11.97 \\
      2016-03-09T13:31 & 1529 & 60084 & 55 & 150 & 10.510 & 9.519 & r & 1.42 & 0.87 & 0.326 & 0.60 & 22.02$\pm$0.03 & 12.02 \\
      2016-03-09T13:34 & 1529 & 60086 & 61 & 150 & 10.510 & 9.519 & r & 1.44 & 0.90 & -- & -- & -- & -- &  \\
      \hline
  \end{tabular}}\label{tab:1stlist}
\begin{tabnote}
\footnotemark[$a$] Observation date and time in UTC\\ 
\footnotemark[$b$] An ID uniquely assigned to the date of HSC observations\\ 
\footnotemark[$c$] An ID uniquely assigned to each exposure\\ 
\footnotemark[$d$] CCD identifier\\ 
\footnotemark[$e$] Exposure time\\ 
\footnotemark[$f$] Heliocentric distance of the comet at observation\\ 
\footnotemark[$g$] Geocentric distance of the comet at observation\\ 
\footnotemark[$h$] Phase angle (Sun-comet-observer angle) at observation\\ 
\footnotemark[$i$] Full width at half maximum of the comet image\\ 
\footnotemark[$j$] Observed apparent magnitudes of the comet in the specified filter ($g$, $r$, or $y$) \\ 
\footnotemark[$k$] Reduced magnitudes at the time of observation\\ 
\end{tabnote}
\end{table}

\end{landscape}

\begin{table}
  \tbl{Derived magnitudes of 28P.}{%
  \begin{tabular}{cccccc}
      \hline
      filter & ObsDate & $\alpha_\mathrm{avg}$\footnotemark[$a$] & $M_{x}\pm\sigma$\footnotemark[$b$] & $H_{x}$\footnotemark[$c$] & $m_{x}(1,1,0.334)$\footnotemark[$d$] \\
             &         & [degrees]        & [mag]                     &       & [mag]    \\                
      \hline
      $y$      & 2016-02-12 & 2.781       & $21.85\pm0.07$ & $11.73\pm0.07$ & $11.74\pm0.07$ \\
      $g$      & 2016-03-07 & 0.521       & $22.82\pm0.06$ & $12.81\pm0.06$ & $12.82\pm0.06$ \\
      $r$      & 2016-03-09 & 0.334       & $22.15\pm0.08$ & $12.14\pm0.08$ & $12.15\pm0.08$ \\
      \hline
    \end{tabular}}\label{tab:mags}
\begin{tabnote}
\footnotemark[$a$] Average phase angles of observations for each band \\ 
\footnotemark[$b$] Mean magnitudes of the comet. $x$ means the filter bands: $g$, $r$, or $y$ \\ 
\footnotemark[$c$] Absolute magnitudes \\ 
\footnotemark[$d$] Reduced magnitudes at the phase angle of \timeform{0.334D} \\ 
\end{tabnote}
\end{table}

\begin{ack}

The Hyper Suprime-Cam (HSC) collaboration includes the astronomical communities of Japan and Taiwan, and Princeton University. The HSC instrumentation and software were developed by the National Astronomical Observatory of Japan (NAOJ), the Kavli Institute for the Physics and Mathematics of the Universe (Kavli IPMU), the University of Tokyo, the High Energy Accelerator Research Organization (KEK), the Academia Sinica Institute for Astronomy and Astrophysics in Taiwan (ASIAA), and Princeton University. Funding was contributed by the FIRST program from the Japanese Cabinet Office, the Ministry of Education, Culture, Sports, Science and Technology (MEXT), the Japan Society for the Promotion of Science (JSPS), Japan Science and Technology Agency (JST), the Toray Science Foundation, NAOJ, Kavli IPMU, KEK, ASIAA, and Princeton University. 

This paper makes use of software developed for Vera C. Rubin Observatory. We thank the Rubin Observatory for making their code available as free software at \url{http://pipelines.lsst.io/}.

This paper is based on data collected at the Subaru Telescope and retrieved from the HSC data archive system, which is operated by the Subaru Telescope and Astronomy Data Center (ADC) at NAOJ. Data analysis was in part carried out with the cooperation of Center for Computational Astrophysics (CfCA), NAOJ. We are honored and grateful for the opportunity of observing the Universe from Maunakea, which has the cultural, historical and natural significance in Hawaii. 

\end{ack}

\section*{Funding}
This work was supported by JSPS KAKENHI Grant Numbers JP23H01234, JP23K25930, JP24H00271, JP23H01217, and JP23K25913

\section*{Data availability} 
 The data underlying this article are available on a dedicated website of HSC-SSP Public Data Release
 (https://hsc.mtk.nao.ac.jp/ssp/data-release/).

\appendix 
\section*{Transformations between HSC magnitudes and $BVR_{C}I_{C}$ magnitude in AB system.}

To convert HSC magnitudes ($m_x$, where $x$ represents the HSC filter band: $g$, $r$, $i$, $z$, and $y$)
to Johnson-Kron-Cousins magnitudes in the AB system, the following equations can be used:

\begin{equation}
  \begin{split}
	B &= m_{\rm g} + 0.0529 \\
	  &\quad + 0.3653 (g - r)_{\rm HSC} + 0.0434 ((g - r)_{\rm HSC})^{2} \\
  \end{split}
\label{B_AB}
\end{equation}

\begin{equation}
  \begin{split}
  V &= m_{\rm g} - 0.0140 \\
    &\quad - 0.5689 (g - r)_{\rm HSC} - 0.0050 ((g - r)_{\rm HSC})^{2} \\
  \end{split}
\label{V_AB}
\end{equation}

\begin{equation}
  \begin{split}
  R_{\rm c} &= m_{\rm r} + 0.0192 \\
            &\quad - 0.1346 (g - r)_{\rm HSC} + 0.0113 ((g - r)_{\rm HSC})^{2} \\
  \end{split}
\label{Rc_AB_1}
\end{equation}

\begin{equation}
  \begin{split}
  R_{\rm c} &= m_{\rm r} - 0.0054 \\
            &\quad - 0.2393 (r - i)_{\rm HSC} - 0.0616 ((r - i)_{\rm HSC})^{2} \\
  \end{split}
\label{Rc_AB_2}
\end{equation}

\begin{equation}
  \begin{split}
  I_{\rm c} &= m_{\rm i} + 0.0128 \\
            &\quad - 0.3849 (i - z)_{\rm HSC} + 0.0441 ((i - z)_{\rm HSC})^{2} \\
  \end{split}
\label{Ic_AB}
\end{equation}

The relationships between AB magnitude systems and Vega magnitude systems are shown below \citep{Fukugita1995}.
\begin{eqnarray}
B_{\rm AB} - B_{\rm Vega} =  -0.110\\
V_{\rm AB} - V_{\rm Vega} =  0.011\\
{R_{\rm c}}_{\rm AB} - {R_{\rm c}}_{\rm Vega} =  0.199\\
{I_{\rm c}}_{\rm AB} - {I_{\rm c}}_{\rm Vega} =  0.456
\end{eqnarray}


\begin{thebibliography}{}

\bibitem[Aihara et al.(2018)]{Aihara2018} Aihara, H., et al.\ 2018, \pasj, 70, S4



\bibitem[Aihara et al.(2022)]{AiharaPDR3} Aihara, H., et al.\ 2022, \pasj, 74, 247



\bibitem[Bach et al.(2019)]{Bach2019} Bach, Y.~P., Ishiguro, M., Jin, S., et al.\ 2019, Journal of Korean Astronomical Society, 52, 71

\bibitem[Belskaya \& Shevchenko(2000)]{Belskaya2000} Belskaya, I.~N., \& Shevchenko, V.~G.\ 2000, Icarus, 147, 94

\bibitem[Belskaya et al.(2002)]{Belskaya2002} Belskaya, I.~N., Shevchenko, V.~G., Efimov, Y.~S., et al.\ 2002, Asteroids, Comets, and Meteors: ACM 2002, 500, 489. 

\bibitem[Bowell et al.(1989)]{Bowell1989} Bowell, E., Hapke, B., Domingue, D., Lumme, K., Peltoniemi, J., \& Harris, A.~W.\ 1989, Asteroids II, 524

\bibitem[Bosch et al.(2018)]{Bosch2018} Bosch, J., et al.\ 2018, \pasj, 70, S5

\bibitem[Bosch et al.(2019)]{Bosch2019} Bosch, J., et al.\ 2019, Astronomical Data Analysis Software and Systems XXVIII, 523, 521

\bibitem[Campins et al.(1987)]{Campins1987} Campins, H., A'Hearn, M.~F., \& McFadden, L.-A.\ 1987, \apj, 316, 847


\bibitem[Campins et al.(2007)]{Campins2007} Campins, H., Licandro, J., Pinilla-Alonso, N., et al.\ 2007, \aj, 134, 1626


\bibitem[Ciarniello et al.(2015)]{Ciarniello2015} Ciarniello, M., Capaccioni, F., Filacchione, G., et al.\ 2015, \aap, 583, A31

\bibitem[Delahodde et al.(2001)]{Delahodde2001} Delahodde, C.~E., Meech, K.~J., Hainaut, O.~R., \& Dotto, E.\ 2001, \aap, 376, 672


\bibitem[Donaldson et al.(2023)]{Donaldson2023} Donaldson, A., Kokotanekova, R., Ro{\.z}ek, A., et al.\ 2023, \mnras, 521, 1518

\bibitem[Fitzsimmons et al.(1994)]{Fitzsimmons1994} Fitzsimmons, A., Dahlgren, M., Lagerkvist, C.-I., Magnusson, P., \& Williams, I.~P.\ 1994, \aap, 282, 634

\bibitem[Fukugita et al.(1995)]{Fukugita1995} Fukugita, M., Shimasaku, K., \& Ichikawa, T.\ 1995, \pasp, 107, 945

\bibitem[Furusawa et al.(2018)]{Furusawa2018} Furusawa, H., Koike, M., Takata, T., et al.\ 2018, \pasj, The on-site quality-assurance system for Hyper Suprime-Cam: OSQAH, 70, S3

\bibitem[Gunn \& Stryker(1983)]{GS1983} Gunn, J.~E. \& Stryker, L.~L.\ 1983, \apjs, 52, 121

\bibitem[Hainaut \& Delsanti(2002)]{Hainaut2002} Hainaut, O.~R., \& Delsanti, A.~C.\ 2002, \aap, 389, 641

\bibitem[Hamilton et al.(2019)]{Hamilton2019} Hamilton, V.~E., Simon, A.~A., Christensen, P.~R., et al.\ 2019, Nature Astronomy, Evidence for widespread hydrated minerals on asteroid (101955) Bennu, 3, 332


\bibitem[Hapke et al.(1998)]{Hapke1998} Hapke, B., Nelson, R., \& Smythe, W.\ 1998, \icarus, 133, 1, 89.

\bibitem[Hapke(2012)]{Hapke2012} Hapke, B.\ 2012, Theory of Reflectance and Emittance Spectroscopy.\ Cambridge: Cambridge University Press

\bibitem[Hasegawa et al.(2014)]{Hasegawa2014} Hasegawa, S., Miyasaka, S., Tokimasa, N., et al.\ 2014, \pasj, 66, 89

\bibitem[Hasselmann et al.(2017)]{Hasselmann2017} Hasselmann, P.~H., Barucci, M.~A., Fornasier, S., et al.\ 2017, \mnras, 469, S550


\bibitem[Ieva et al.(2022)]{Ieva2022} Ieva S., Arcoverde P., Rond{\'o}n E., Giunta A., Dotto E., Lazzaro D., Mazzotta Epifani E., et al., 2022, MNRAS, 513, 3104


\bibitem[Jewitt \& Luu(1990)]{Jewitt1990} Jewitt, D.~C., \& Luu, J.~X.\ 1990, \aj, 100, 933

\bibitem[Jewitt \& Meech(1988)]{Jewitt1988} Jewitt, D.~C., \& Meech, K.~J.\ 1988, \apj, 328, 974

\bibitem[Kitazato et al.(2019)]{Kitazato2019} Kitazato, K., Milliken, R.~E., Iwata, T., et al.\ 2019, Science, 364, 272


\bibitem[Komiyama et al.(2018)]{Komiyama2018} Komiyama, Y., et al.\ 2018, \pasj, 70, S2

\bibitem[Lamy et al.(2004)]{Lamy2004} Lamy, P.~L., Toth, I., Fernandez, Y.~R., et al.\ 2004, Comets II, 223

\bibitem[Lamy et al.(2011)]{Lamy2011} Lamy, P.~L., Toth, I., Weaver, H.~A., et al.\ 2011, \mnras, 412, 3, 1573


\bibitem[Mann \& von Braun(2015)]{Mann2015} Mann, A.~W. \& von Braun, K.\ 2015, \pasp, 127, 948, 102

\bibitem[Masoumzadeh et al.(2017)]{Masoumzadeh2017} Masoumzadeh, N., et al.\ 2017, \aap, 599, A11

\bibitem[Masoumzadeh et al.(2019)]{Masoumzadeh2019} Masoumzadeh, N., et al.\ 2019, \aap, 630, A11

\bibitem[Meech et al.(2004)]{Meech2004} Meech, K.~J., Hainaut, O.~R., \& Marsden, B.~G.\ 2004, Icarus, 170, 2, 463


\bibitem[Miyazaki et al.(2018)]{Miyazaki2018} Miyazaki, S., et al.\ 2018, \pasj, 70, S1

\bibitem[Muinonen et al.(2010)]{Muinonen2010} Muinonen, K., Belskaya, I.~N., Cellino, A., Delb{\`o}, M., Levasseur-Regourd, A.-C., Penttil{\"a}, A., \& Tedesco, E.~F.\ 2010, Icarus, 209, 542



\bibitem[Ootsubo et al.(2021)]{Ootsubo2021} Ootsubo, T., Kawakita, H., \& Shinnaka, Y.\ 2021, Icarus, 363, 114425

\bibitem[Ootsubo et al.(2025)]{Ootsubo2025} Ootsubo, T., Takata, T., Furusawa, J., et al.\ 2025, Astronomical Data Analysis Software and Systems XXXII, 538, 397

\bibitem[Penttil{\"a} et al.(2016)]{Penttila2016} Penttil{\"a}, A., Shevchenko, V.~G., Wilkman, O., \& Muinonen, K.\ 2016, \planss, 123, 117

\bibitem[Rivkin et al.(2002)]{Rivkin2002} Rivkin, A.~S., Howell, E.~S., Vilas, F., et al.\ 2002, Asteroids III, 235

\bibitem[Rosenbush et al.(2002)]{Rosenbush2002} Rosenbush, V., Kiselev, N., Avramchuk, V., et al.\ 2002, in Optics of Cosmic Dust, ed. G. Videen \& M. Kocifaj (Dordrecht: Kluwer Academic Publishers), 191. 

\bibitem[Sakugawa et al.(2018)]{Sakugawa2018} Sakugawa H., Terai T., Ohtsuki K., Yoshida F., Takato N., Lykawka P.~S., Wang S.-Y., 2018, PASJ, 70, 116

\bibitem[Schleicher et al.(2022)]{Schleicher2022} Schleicher, D., Knight, M., Skiff, B., \& Bair, A.\ 2022, AAS/Division for Planetary Sciences Meeting Abstracts, 54, 309.03


\bibitem[Shevchenko(1996)]{Shevchenko1996} Shevchenko, V.~G.\ 1996, Lunar and Planetary Science Conference, 27, 1193

\bibitem[Shevchenko et al.(2016)]{Shevchenko2016} Shevchenko, V.~G., Belskaya, I.~N., Muinonen, K., et al.\ 2016, \planss, 123, 101. doi:10.1016/j.pss.2015.11.007


\bibitem[Terai et al.(2018)]{Terai2018} Terai, T., et al.\ 2018, \pasj, 70, S40


\bibitem[Usui et al.(2019)]{Usui2019} Usui, F., Hasegawa, S., Ootsubo, T., et al.\ 2019, \pasj, 71, 1








\bibitem[Willmer(2018)]{Willmer2018} Willmer, C.~N.~A.\ 2018, \apjs, 236, 47

\end{thebibliography}

\end{document}